\documentclass[doublecol]{epl2} 
\usepackage{subcaption}
\usepackage{dcolumn}
\usepackage{booktabs}
\usepackage{adjustbox}
\usepackage{amsmath}

\title{First-principles phonon calculations for lattice dynamics, thermal expansion and lattice thermal conductivity of CoSi at high temperature region}

\shorttitle{Title} 

\author{Shamim Sk\inst{1\footnote{E-mail: shamimsk20@gmail.com}} \and Sudhir K. Pandey\inst{2\footnote{E-mail: sudhir@iitmandi.ac.in}}}
\shortauthor{S. Sk \etal}

\institute{                    
  \inst{1} School of Basic Sciences, Indian Institute of Technology Mandi, Kamand - 175075, India\\
  \inst{2} School of Engineering, Indian Institute of Technology Mandi, Kamand - 175075, India
}

\abstract{This study presents the first-principles phonon calculations to understand the experimental thermal expansion ($\alpha(T)$) and lattice thermal conductivity ($\kappa_{L}$) of CoSi at high temperature region. Phonon dispersion is computed using finite displacement method and supercell approach by taking the equilibrium crystal structures obtained from DFT. The calculation of $\alpha(T)$ is done under quasi-harmonic approximation. The $\kappa_{L}$ is calculated using first-principle anharmonic lattice dynamics calculations under single-mode relaxation time approximation. Calculated $\alpha(T)$ in the temperature range $0-1300$ K gives the good match with existing experimental data. The calculated value of $\kappa_{L}$ ($\sim$8.0 W/m-K) at 300 K is found to be in good agreement with the experimental value of $\sim$8.3 W/m-K. The temperature dependent of phonon lifetime due to phonon-phonon interaction is calculated to understand the behaviour of $\kappa_{L}$. Present study suggests that ground state phonon dispersion obtained from DFT based methods gives reasonably good explanation of experimental $\alpha(T)$ and $\kappa_{L}$.}

\begin{document}

\maketitle

\section{Introduction}
Theoretical understanding of the thermal transport properties by phonon is a challenging task in condensed matter physics and materials science. The source of scattering mechanisms for realizing the phonon transport properties are phonon-phonon interaction, electron-phonon interaction, phonon-defect interaction etc\cite{ashcroft}. The capturing of these scattering mechanisms for any system is really a challenging job computationally due to involvement of the many-body interactions. At finite temperature, the study of these properties becomes more difficult at the level of theory and computation due to the presence of anharmonic effect. However, the recent advancement of high performance computers improves the situation to some extent. For instance, to study the lattice dynamics and dependent phonon properties at finite temperature, an \textit{ab initio} molecular dynamics simulations is recently used\cite{lattice1,lattice2,lattice3,lattice4}. But, this method is computationally costly and the implementation of this method is not straightforward. In this context, the reliable and relatively cheaper methods to understand the phononic properties are first-principle density functional theory (DFT)\cite{dft} based techniques. 

In addition to the electronic structure information, DFT also calculates the force exerted on each atom of the compound. Once atom is displaced from its equilibrium position, the forces of all atoms enhance. The systematic displacement of atoms give a number of phonon frequencies. The method for analysis of these phonon frequencies known as finite displacement method (FDM)\cite{fdm}. The another method to analysis the phonon frequencies is density functional perturbation theory (DFPT)\cite{dfpt}. At the DFT level, it is the common practice to use the ground state phonon dispersion to calculate the phonon properties at finite temperature. But, the ground state of DFT itself faces many challenges. For instance, many-particle wave function is approximated by one-particle wave function in DFT. Apart from this, DFT approximates the ground state results by various exchange-correlation (XC) functionals. On the top of these challenges in ground state, it will be interesting to study up to what extent DFT addresses the phonon properties at high temperature. Keeping such a challenge in mind, we have chosen CoSi as a case example to study the phonon related properties at high temperature region.    

CoSi is recently marked as novel topological semimetal\cite{tang,takane}. This compound with a B20 simple cubic structure is reported as a potential thermoelctric (TE) material from last few decades\cite{asanable,kim,lue,ren,li,kuo,pan,skoug,sun_2013,sun_2017,longhin,yu}. The TE material is one which converts the waste heat into useful electricity. The efficiency of TE materials is calculated using the dimensionless parameter, called figure of merit\cite{zt}, $ZT=S^{2}\sigma T/\kappa$. Where, S is Seebeck coefficient, $\sigma$ is electrical conductivity, $\kappa$ is thermal conductivity and T is absolute temperature of the material. The $\kappa$ consists of two parts: electronic thermal conductivity ($\kappa_{e}$) and lattice thermal conductivity ($\kappa_{L}$). The efficient TE materials should have the value of \textit{ZT} greater or equal to unity\cite{snyder}. Therefore, potential TE materials should possess high power factor (PF = $S^{2}\sigma$) with low $\kappa$. The PF of CoSi is comparable with so called state-of-the-art TE materials Bi$_{2}$Te$_{3}$\cite{bite} and PbTe\cite{pbte}. But, \textit{ZT} of CoSi is diminished by it's high $\kappa$. Actually, achieving high \textit{ZT} is really a difficult task, as S, $\sigma$ and $\kappa_{e}$ are strongly dependent to each other through charge carrier\cite{ashcroft,shamim_mrx}. Minimizing of $\kappa_{e}$ without affecting $\sigma$ is a difficult job as they are related linearly via Wiedeman-Franz law: $\kappa_{e}=L\sigma T$, \textit{L} is Lorenz number. Hence, the brilliant way to maximize \textit{ZT} of CoSi is to optimize $\kappa_{L}$. Therefore, it is necessary to study the phonon properties in order to understand $\kappa_{L}$. 

In the context of thermoelectricity, phonon plays a crucial role to fully evaluate the TE materials. The information of phonon must be considered for understanding the lattice thermal conductivity, thermal expansion etc. Here, the ground state phonon dispersion obtained from DFT based method can be used to study the temperature dependent phononic transport properties under various approximations. Within the harmonic and quasi-harmonic approximations the basic phonon properties e.g. phonon dispersion, heat capacity at constant volume (C$_{v}$), entropy, thermal expansion, heat capacity at constant pressure (C$_{p}$) etc. can be calculated. But, calculating the lattice thermal conductivity needs the anharmonic force constant which requires the many-body perturbation theory and hence computational implementation becomes more challenging as well as time consuming. In addition to the different approximations involved in calculating the phonon properties, the different XC functionals, force constant cutoff, supercell size, atomic displacement size etc. strictly affect the ground state phonon dispersion\cite{lattice1}. Therefore, by taking this ground state phonon dispersion, it will be a great challenge to understand the experimental phonon related properties at high temperature.                        

In this work, the first-principle DFT based phonon calculations are carried out to understand the experimental $\alpha(T)$ and $\kappa_{L}$ at high temperature region. The finite displacement method and supercell approach are used to calculate the phonon dispersion. The calculated thermal expansion in the temperature range $0-1300$ K gives the nice match with the experimental reported data. The anharmonic lattice dynamics has been introduced to calculate $\kappa_{L}$. The computed value of $\kappa_{L}$ is $\sim$8.0 W/m-K which is in consistent with the estimated experimental value of $\sim$8.3 W/m-K at room temperature. Temperature dependent phonon lifetime due to phonon-phonon interaction is calculated in order to understand the experimental feature of $\kappa_{L}$.

\section{COMPUTATIONAL DETAILS}
Phonon properties are calculated using PHONOPY code\cite{phonopy} based on finite displacement method (FDM) and supercell approach\cite{fdm}. A supercell of size 2 $\times$ 2 $\times$ 2 containing 64 atoms is used in order to calculate the total forces on each atom in WIEN2k code\cite{wien2k}. The local density approximation (LDA)\cite{lda} is used as XC functional. The k-mesh size of 4 $\times$ 4 $\times$ 4 is used in the full Brillouin zone for force calculation. The convergence criteria for the calculations of forces is set as 0.1 mRy/Bohr. Using these forces, second order force constants are extracted using PHONOPY code to calculate the phonon dispersion. Thermal expansion of this compound is also calculated under quasi-harmonic approximation (QHA) as implemented in PHONOPY code. 

The lattice thermal conductivity is calculated using PHONO3PY code\cite{phono3py} within supercell approach. Same size of supercell is used here as used in PHONOPY. The ABINIT software\cite{abinit} is employed to calculate forces on each atom using projector augmented wave (PAW) method under DFT. The PAW datasets are taken from the work of Jollet \textit{et. al}\cite{jollet}. The LDA\cite{lda} is employed as XC functional which was also used in the WIEN2k force calculation in the case of PHONOPY. The self-consistency is achieved by setting the force convergence criteria of 5 $\times$ 10$^{-8}$ Ha/Bohr. The k-mesh size of 4 $\times$ 4 $\times$ 4 is used in the full Brillouin zone.	

The output results of ABINIT containing forces on each atom are used in PHONO3PY\cite{phono3py} to calculate second and third order force constants. These force constants are used to calculate lattice thermal conductivity using single-mode relaxation time
approximation as implemented in PHONO3PY. A dense q-mesh size of 21 $\times$ 21 $\times$ 21 is used for lattice thermal conductivity calculation. A real-space cutoff distance of 5.16 Bohr is set to ensure the three neighbour atoms interaction. This setting distance effectively reduces the number of supercell calculations and hence minimizes the computational cost.

\section{RESULTS AND DISCUSSION}      
Phonon dispersion of CoSi is presented in fig. 1 along the high symmetry directions in the first Brillouin zone. The positive phonon frequencies (or energies) of all the branches signifying the mechanical stability of the compound\cite{phonopy}. Phonon dispersion associates with twenty four branches as primitive cell of CoSi contains eight atoms. Out of them three are acoustic branches and twenty one are optical branches. The maximum phonon energy is calculated as $\sim$52 meV, which closely matches with the experimental value of 56 meV\cite{delaire}. At $\sim$22 mev, all the three optical branches are nearly degenerated with the five optical branches at $R$ point. The nature of acoustic branches along $\Gamma-M$ and $\Gamma-R$ directions are almost linear, which means that the group velocity is nearly close to phase velocity at this region\cite{ashcroft}. Here, it is important to note that the sound velocity can be calculated from the slopes of acoustic branches\cite{ashcroft}. This sound velocity is associated with the lattice thermal conductivity in a solid\cite{ashcroft,macia}.

\begin{figure}
\begin{center}
\includegraphics[width=0.7\linewidth, height=5.0cm]{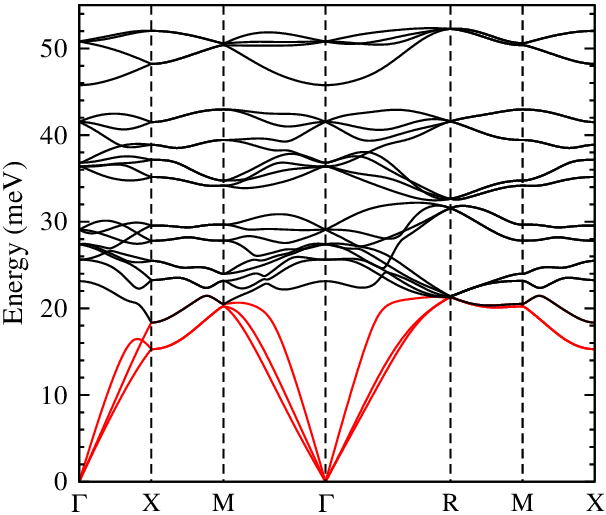} 
\caption{\small{Phonon dispersion of CoSi.}}
\end{center}
\end{figure}

A detailed study of the dependency of XC functionals on vibrational and thermodynamic properties of CoSi has been carried out by Sk \textit{et al}\cite{shamim_physicab}. Though the vibrational properties were reported as XC functional dependent, but thermodynamic properties were almost insensitive of XC functional. However, LDA was reported to give the better value of Debye temperature as compared to other functionals. Here, we have chosen LDA as an XC functional for calculating all the phonon related properties in this work. 

In a pure harmonic solid, the phonon related properties like thermal expansion and lattice thermal conductivity are not defined as phonons do not interact under harmonic approximation\cite{lindsay}. In this context, quasi harmonic approximation is found to be a reasonably good approximation to capture the thermal expansion. The study of thermal expansion is much needful in those areas where temperature dependent properties of the materials  are taken into account. The materials used for making modules of TEG experience temperature gradient. Hence, study of thermal expansion of these materials are helpful before putting these materials in TEG. Keeping this in mind, we have calculated linear thermal expansion coefficient ($\alpha (T)$) of CoSi under QHA. Here, QHA used for volume dependence of phonon properties. fig. 2(c) displays the calculated $\alpha (T)$ in the temperature range $0-1300$ K. For the better understanding of fig. 2(c), it is necessary to explain fig. 2(a) and (b) first. Fig. 2(a) exhibits the change in total free energy as a function of primitive cell volume at different temperatures starting from 0 K to 1300 K with step size of 100 K. Here, the total free energy at given temperature and volume is expressed as: $F(T; V) = [U_{el}(V)-U_{el}(V_{0})]+F_{ph}(T; V)$. Where $U_{el}(V)-U_{el}(V_{0})$ is the relative ground state electronic energy obtained from first-principle calculation, $V_{0}$ is the equilibrium volume at 0 K. $F_{ph}(T; V)$ is the phonon Helmholtz free energy. Fig. 2(a) shows that for every temperature there are energy minima (denoted by solid square) corresponding to equilibrium volume of primitive cell which are connected by solid line (red). These equilibrium volumes are plotted as a function of temperature as presented in fig. 2(b). The primitive cell volume increases monotonically with increase in temperature after 100 K. The equilibrium volume at 0 K is calculated as $\sim$83.6 \AA$^{3}$, whereas it reaches $\sim$86.4 \AA$^{3}$ at 1300 K. The volume is increased by $\sim$3.3$\%$ in the temperature interval $0-1300$ K. 

\begin{figure}
\begin{center}
\includegraphics[width=0.9\linewidth, height=6.0cm]{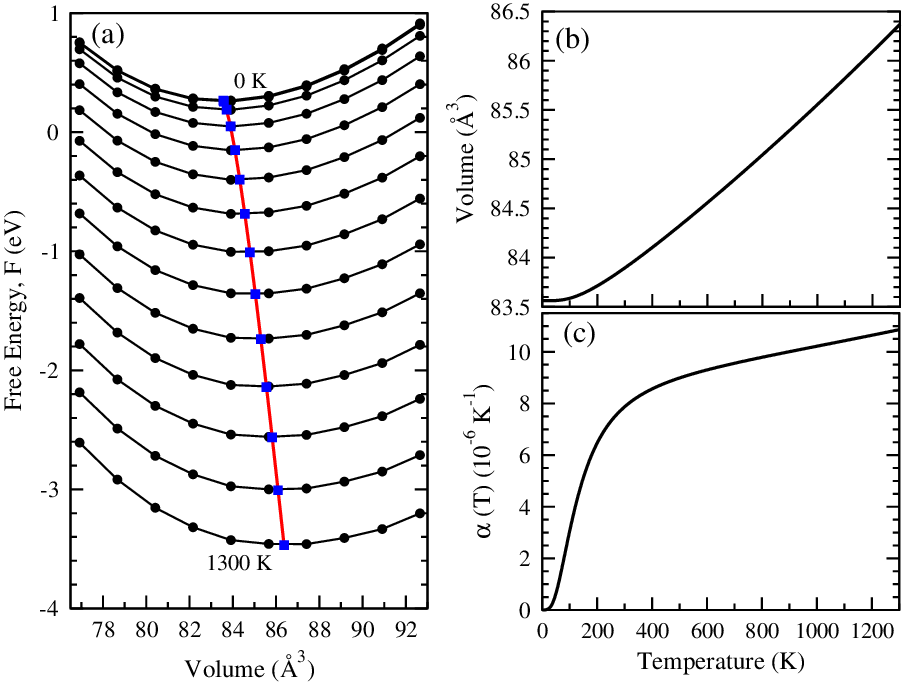} 
\caption{\small{(a) Variation of total free energy $F$ with primitive cell volume. (b) Change in primitive cell volume with temperature. (c) Linear thermal expansion coefficient $\alpha (T)$ as a function of temperature for CoSi.}}
\end{center}
\end{figure}

\begin{table*}
\begin{center}
\caption{\label{tab:table1}%
\small{Calculated linear thermal expansion coefficients ($\alpha (T)$) at different temperatures compared with reported experimental values.}}
\setlength{\tabcolsep}{15pt}
\begin{tabular}{lcccccc}
\toprule
\toprule
& & & & \textrm{$\alpha (T)$ [$\times10^{-6}$K$^{-1}$]}\\
\toprule
\textrm{Reported}&
\textrm{100 K}&
\textrm{300 K}&
\textrm{600 K}&
\textrm{900 K}&
\textrm{1200 K}&
\textrm{1300 K}\\ 

\toprule
This work                                    & 3.1 & 7.9 & 9.3 & 10.0 & 10.6 & 10.9 \\
Mandrus \textit{et al.\cite{mandrus}}       & 3.8 & 9.8 & --- & --- & --- & --- \\
Krentsis \textit{et al.\cite{krentsis}}       & 3.9 & 10.4 & 11.5 & 12.1 & --- & --- \\
Ruan \textit{et al.\cite{ruan}}             & --- & --- & 12.4 & 13.4 & 15.2 & 15.8 \\
\toprule
\toprule
\end{tabular}
\end{center}
\end{table*}

Using the primitive cell volume at different temperatures (as shown in fig. 2(b)), we have calculated the volumetric thermal expansion coefficient as follows: $\beta(T) = \frac{1}{V(T)}\frac{\partial V(T)}{\partial T}$. CoSi has simple cubic structure, hence considering uniform expansion in all the three directions, $\alpha (T)$ can be taken as one third of $\beta(T)$\cite{ashcroft}. The calculated $\alpha (T)$ is plotted in fig. 2(c). The same method to calculate $\alpha (T)$ is used by earlier reported work\cite{shastri_4th,shastri_5th}. Fig. 2(c) shows the rapid increment of $\alpha (T)$ up to $\sim$300 K, then increases slowly till 1300 K. The average increment rate of $\alpha (T)$ in the temperature range $0-300$ K is calculated as $\sim$0.03/K, whereas this value is observed as $\sim$0.004/K in the temperature range $300-1300$ K. The values of $\alpha (T)$ at 300 K and 1300 K are found to be $\sim$7.9$\times$10$^{-6}$ K$^{-1}$ and $\sim$10.9$\times$10$^{-6}$ K$^{-1}$, respectively. The calculated values of $\alpha (T)$ are compared with experimental reported data (dilatometer measurement) as displayed in Table 1. Table shows that at low temperature calculated $\alpha (T)$ gives good match with experiment, but as the temperature increases calculated $\alpha (T)$ deviates slightly from experiment with lower values. However, the values of $\alpha (T)$ calculated under QHA are found in good agreement with the reported data.

The study of $\alpha (T)$ gives the change in length of TE materials when subjected to a heating or cooling cycle. In many cases, the TE materials need to face the enough mechanical stress during a large number of heating and cooling cycles, which is named as thermal fatigue. In such cases, study of thermal fatigue is much helpful before putting the TE materials in TEG for long time application of TEG. The product of elastic modulus and $\alpha (T)$ is a useful quantity to study the thermal fatigue\cite{case,music}. The Calculated alpha(T) can be used in designing TEG. Thermal expansion is also responsible for the microcracking and porosity in TE materials, which affect the performance of TE materials. For instance, Zhang \textit{et al.}\cite{zhang} studied the effect of microcracking on TE properties of skutterudite specimen in the temperature range $300-800$ K. They showed that presence of microcracks affect the Seebeck coefficient in minor scale but they observed drastic change in electrical conductivity. Subsequently, power factor decreases and gives low \textit{ZT} as they said\cite{zhang}. Therefore, before using of any TE material for making TEG, study of $\alpha (T)$ is much helpful for long time application of TEG.

The lattice part of thermal conductivity ($\kappa_{L}$) is calculated using first-principle anharmonic lattice dynamics calculations implemented in PHONO3PY code\cite{phono3py}. Anharmonic force constants are computed from first-principle calculation. First, phonon lifetime is calculated from the imaginary part of phonon self-energy by considering the phonon-phonon interaction only. The third order force constants contain anharmonicity are employed for calculating the imaginary part of self-energy. The lifetime of the phonon mode $\lambda$ is given by\cite{maradudin,togo_2015}
\begin{equation}
\tau_{\lambda}=\frac{1}{2\Gamma_{\lambda}(\omega_{\lambda})},
\end{equation}
where, $2\Gamma_{\lambda}(\omega_{\lambda})$ corresponds to the phonon linewidth of the phonon mode $\lambda$ and $\omega_{\lambda}$ is the frequency of a phonon mode. This phonon lifetime is related with $\kappa_{L}$. The $\kappa_{L}$ is computed by solving the linearized phonon Boltzmann equation (LBTE) under single-mode relaxation time (SMRT) method and given by\cite{togo_2015,srivastava}
\begin{equation}
\kappa_{L}=\frac{1}{NV_{0}}\sum_{\lambda}C_{\lambda}\textbf{v}_{\lambda}\otimes \textbf{v}_{\lambda}\tau_{\lambda}^{SMRT},
\end{equation}
where, \textit{N} is the number of unit cells in the crystal, $V_{0}$ is volume of a unit cell, $\textbf{v}_{\lambda}$ and $\tau_{\lambda}^{SMRT}$ are the group velocity and single-mode relaxation time of phonon mode $\lambda$. $C_{\lambda}$ is the mode dependent heat capacity. For calculating $\kappa_{L}$, the single-mode relaxation time $\tau_{\lambda}^{SMRT}$ is considered as the phonon lifetime $\tau_{\lambda}$, where $\tau_{\lambda}$ is defined in eq. 1.

\begin{figure}
\begin{center}
\includegraphics[width=0.7\linewidth, height=5.5cm]{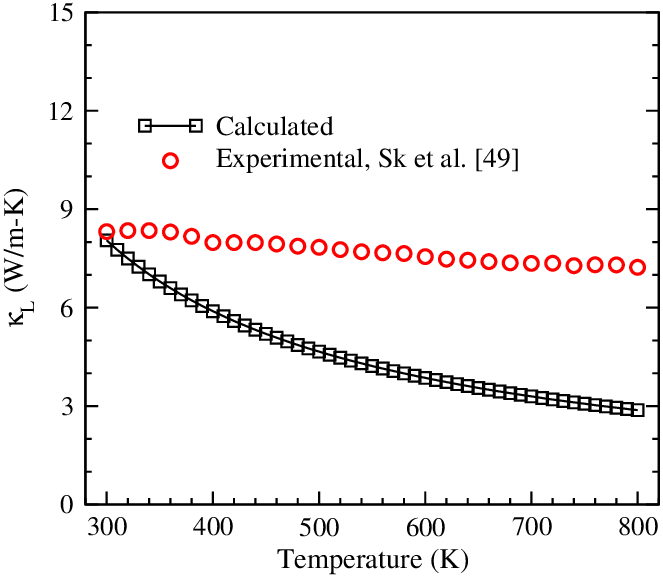} 
\caption{\small{Lattice thermal conductivity as a function of temperature.}}
\end{center}
\end{figure}

The calculated values of $\kappa_{L}$ for CoSi is shown in fig. 3 in the temperature range $300-800$ K. The calculated $\kappa_{L}$ is compared with the experiment in the same figure. Experimental $\kappa_{L}$ is taken from the work of Sk \textit{et al.}\cite{shamim_arxiv} by subtracting the experimental $\kappa_{e}$ (estimated from Wiedeman-Franz law) from the total experimental $\kappa$. The $\kappa_{L}$ at 300 K is calculated as $\sim$8.0 W/m-K which is closely matching with the experimental value of $\sim$8.3 W/m-K at the same temperature. As the temperature increases, the calculated values deviate from the experimental value. At 800 K the calculated and experimental values are found to be $\sim$2.9 W/m-K and $\sim$7.2 W/m-K, respectively. This deviation may be due to the various factors. For instance, Lorenz number, \textit{L} is taken as constant for estimating $\kappa_{e}$, but in practical \textit{L} is temperature dependent quantity. For calculating temperature dependence of $\kappa_{L}$, ground state phonon mode was used. But, thermal expansion of fig. 2(b) shows that volume of primitive unitcell of CoSi is changed with temperature, resulting a change in lattice parameter. Hence, phonon dispersion are expected to be changed once we use temperature dependent lattice parameters. This temperature dependent phonon dispersion may improve the $\kappa_{L}$ value. Therefore, considering all these factors into account, one can expect the better matching between calculated and experimental data at high temperature region, which is beyond the scope of our present study.

In order to understand the temperature dependent behaviour of $\kappa_{L}$, we have calculated the phonon lifetime. The lifetime of phonon in a solid is decided by the various scattering mechanisms, e.g. phonon-phonon interaction (PPI), electron-phonon interaction, phonon-defect interaction etc. Here, we consider only PPI to calculate the phonon lifetime. The lifetime of each phonon mode $\lambda$ is calculated using the imaginary part of phonon self-energy as eq. 1. The imaginary part of the self-energy $\Gamma_{\lambda}(\omega_{\lambda})$ is calculated within many-body perturbation theory as\cite{togo_2015}
\begin{eqnarray}
\Gamma_{\lambda}(\omega)=\frac{18\pi}{\hbar^{2}}\sum_{\lambda^{'}\lambda^{''}}|\Phi_{-\lambda\lambda^{'}\lambda^{''}}|^{2}\{(n_{\lambda^{'}}+n_{\lambda^{''}}+1) \nonumber \\
\times \delta(\omega-\omega_{\lambda^{'}}-\omega_{\lambda^{''}})+(n_{\lambda^{'}}-n_{\lambda^{''}}) \nonumber \\
\times [\delta(\omega+\omega_{\lambda^{'}}-\omega_{\lambda^{''}})-\delta(\omega-\omega_{\lambda^{'}}-\omega_{\lambda^{''}})]\},
\end{eqnarray}  
where, $\Phi_{-\lambda\lambda^{'}\lambda^{''}}$ signifies the strength of interaction among three phonons $\lambda$, $\lambda^{'}$ and $\lambda^{''}$ involved in the scattering. $n_{\lambda}$ is the phonon occupation number at the equilibrium.

Fig. 4(a) shows the lifetime of twenty four branches (marked as b1 to b24) in the temperature range $300-800$ K as calculated by taking the weight average over the q-points in the Brilloun zone. The same method to calculate phonon lifetime is employed earlier\cite{shastri_6th}. From fig. 4(a) it is seen that optical branch b8 has the highest lifetime in the full temperature range signifies the lesser scattering as compared to other phonon branches. Two optical branches b23 and b24 (having almost same lifetime) show the lowest lifetime implying the highest scattering. Finally, the acoustic, optical and total phonon lifetimes are calculated by averaging the respective number of phonon branches as shown in fig. 4(b). Figure shows that the acoustic branches have the higher lifetime than optical branches. This suggests that the contribution of lifetime in $\kappa_{L}$ from acoustic branches are higher than the optical branches. The total lifetime of phonons due to PPI is calculated by averaging all the branches. The total phonon lifetime is calculated as $\sim2.40\times10^{-12}$ s at 300 K, then as the temperature increases this value decreases up to $\sim0.85\times10^{-12}$ s at 800 K. This decrement nature of phonon lifetime with temperature is accordance with the temperature dependent trend of $\kappa_{L}$ (see fig. 3). At this point it is important to note that for calculating $\kappa_{L}$ (see eq. 2), group velocity is taken as temperature independent quantity. In the work of Sk \textit{et al.}\cite{shamim_physicab} specific heat of CoSi has been shown to increase with increasing temperature and almost constant at high temperature (above $\sim$500 K). Therefore, the decrement nature of $\kappa_{L}$ is mainly determined by the temperature dependent trend of phonon lifetime. The phonon lifetime can be reduced (in order to maximize \textit{ZT}) by introducing the extra scattering centres via nanostructuring, alloying etc\cite{snyder,djsingh}. In this way $\kappa_{L}$ can be minimized to get high \textit{ZT} of CoSi.        

\begin{figure}
\begin{center}
\includegraphics[width=0.9\linewidth, height=10.0cm]{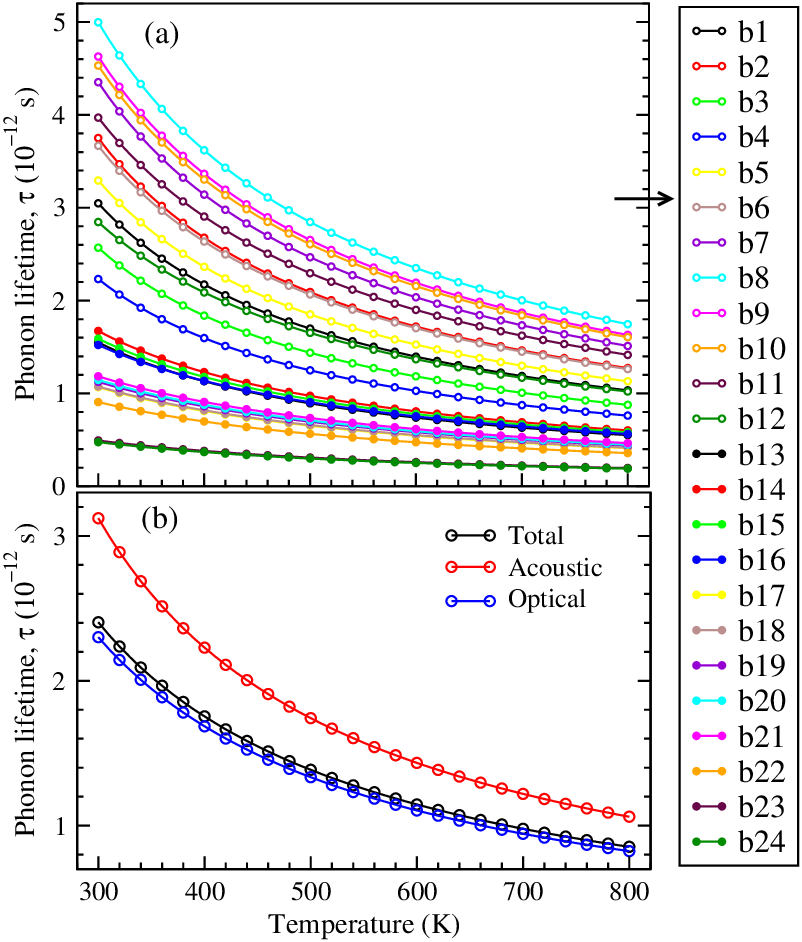} 
\caption{\small{(a) The phonon lifetime for twenty four phonon branches. (b) Total, acoustic, optical phonon lifetime due to PPI as a function of temperature.}}
\end{center}
\end{figure}

\section{CONCLUSIONS}
In summary, we have studied the first-principles DFT based phonon calculations to understand the phonon properties of CoSi at high temperature region. The finite displacement method has been employed to compute the phonon dispersion. Then the ground state phonon dispersion is used to address the experimental $\alpha(T)$) and $\kappa_{L}$. The calculated QHA based thermal expansion gives quite good match with the existing experimental results. The $\kappa_{L}$ is calculated using anharmonic force constant under many-body perturbation theory. The calculated value of $\kappa_{L}$ ($\sim$8.0 W/m-K) is in good agreement with the experimental value ($\sim$8.3 W/m-K) at room temperature. The temperature dependent behaviour of $\kappa_{L}$ is seen to decide by the trend of phonon lifetime. This study suggests that the phonon band-structure obtained from DFT based method addresses the experimental $\alpha(T)$) and $\kappa_{L}$ in reasonably good manner. However, for better quantification of the studied phonon properties specially at high temperature, one may require to go beyond the DFT based methods.

\end{document}